\documentclass[a4paper,11pt]{article}
\usepackage{jheppub}
\pdfoutput=1

\usepackage{cancel}
\usepackage[normalem]{ulem}
\usepackage{float}
\usepackage{autobreak}
\usepackage{graphicx}
\usepackage{appendix}
\usepackage{amsfonts}
\usepackage{comment}
\usepackage{multirow}
\usepackage{slashed}
\usepackage{amsmath}
\usepackage{float}
\usepackage{autobreak}
\usepackage{graphicx}
\usepackage{appendix}
\usepackage{comment}

\usepackage{tikz}
\usepackage{subcaption}
\usetikzlibrary{decorations.pathmorphing,decorations.markings,calc}
\tikzset{
  fermion/.style={thick,postaction={decorate},
    decoration={markings,mark=at position .55 with {\arrow[scale=1.1]{latex}}}},
  gluon/.style={thick,decorate,
    decoration={coil,amplitude=3.2pt,segment length=4.2pt,aspect=0.9}},
  photon/.style={thick,decorate,
    decoration={snake,amplitude=2.2pt,segment length=7pt}},
  zboson/.style={photon},
  vtx/.style={circle,fill,inner sep=1.1pt},
  lab/.style={font=\small},
}
\newcommand{\diagframe}{\path[use as bounding box] (-2.9,-2.3) rectangle (3.3,2.3);}
\newcommand{\diagscale}{0.85}

\usepackage[dvipsnames]{xcolor}
\usepackage{hyperref}
\usepackage[T1]{fontenc}
\usepackage{tocloft}
\usepackage{placeins}
\usepackage[normalem]{ulem}
\usepackage{subcaption}
\usepackage{array}
\usepackage{enumitem}
\usepackage{extarrows}
\usepackage{mathtools}
\usepackage{slashed} 
\usepackage{color}
\usepackage{soul}
\sethlcolor{yellow} 

\newcommand{\nn}{\nonumber\\}

\newcommand{\Ca}{C_A}

\newcommand{\df}{{\rm d}}
\def\Dm1{{{\delta(1-z)}}}

\def\g0#1DY{{g_{0#1}^{DY}}}

\def\LogmW1{{{\ln (1-\omega)}}}

\newcommand{\overbar}[1]{\,\overline{\!{#1}}}
\newcommand{\Nbar}{\overbar{N}}

\newcommand{\as}{a_s}
\newcommand{\muf}{\mu_F}
\newcommand{\mur}{\mu_R}

\newcommand{\Lw}{L_\omega}
\newcommand{\wb}{\overbar{\omega}}

\newcommand{\eq}[1]{Eq.\ (\ref{#1})}

\title{{
$Z\gamma$ production in gluon fusion at NLO+NLL}}
\author[a,b,c]{Chinmoy Dey,}
\emailAdd{chinmoy@sjtu.edu.cn}
\affiliation[a]{Tsung-Dao Lee Institute, School of Physics and Astronomy, \\Shanghai Jiao Tong University, Shanghai 200240, China}
\affiliation[b]{Shanghai Key Laboratory for Particle Physics and Cosmology, \\
Shanghai Jiao Tong University, Shanghai 200240, China}
\affiliation[c]{Key Laboratory for Particle Astrophysics and Cosmology (MOE), \\
Shanghai Jiao Tong University, Shanghai 200240, China}
\author[d]{Niraj Koirala,}
\emailAdd{n.koirala@iitg.ac.in}
\affiliation[d]{Department of Physics, Indian Institute of Technology Guwahati,\\ Guwahati-781039, Assam, India}
\author[d]{M. C. Kumar,}
\emailAdd{mckumar@iitg.ac.in}
\author[d]{and Vaibhav Pandey}
\emailAdd{vphiitg@iitg.ac.in}

\abstract{ 
We study the production of $Z\gamma$
in gluon fusion 
at hadron colliders 
at the next-to-leading-order (NLO) in QCD.
At the $13.6$ TeV LHC, we find that these NLO corrections are found to enhance the leading-order (LO) cross section by about $83$\% and reduce the conventional seven-point scale uncertainty from $22.3$\% at LO to $15.2$\% at NLO.
We further perform the threshold resummation at next-to-leading-logarithmic (NLL) accuracy and match to the fixed order NLO results. 
We have presented  the production
cross section for various center of mass energies at the LHC. We find that this NLL resummation enhances the NLO production cross section by about $23$\%.
We also give  the invariant mass distribution up to $Q=1000$ GeV for $13.6$ TeV LHC. In this case, the resummation  reduces the seven-point scale uncertainty from
$25\%$ at NLO to about $16$\% at NLO+NLL for $Q=1000$ GeV. We also present the uncertainties due to the strong coupling constant and the parton distribution functions in our results. We further estimate the contribution of the top quark in our predictions, using the Born Improved Theory (BIT).}

\begin{document}

\raggedbottom

\maketitle

\raggedbottom

\tableofcontents


\section{Introduction}
The production of a $Z$ boson in association with a photon is an important channel for precision studies of the electroweak sector at hadron colliders. 
The process has a clean, experimentally identifiable signature in leptonic decay modes of the $Z$-boson and probes the interactions of neutral gauge bosons. 
In the Standard Model, the $ZZ \gamma$ and $Z \gamma \gamma$ vertices are absent at tree level. 
Consequently, measurements of $Z \gamma$ production, particularly in kinematic regions with large photon transverse momentum or large $Z \gamma$ invariant mass, provide sensitivity to possible deviations from the Standard Model predictions and are used to constrain anomalous neutral triple-gauge couplings. 
Recent LHC measurements~\cite{CMS:2026yaq} have examined both fiducial rates and differential distributions, using the latter to test Standard Model predictions and set limits on such couplings. 
The $Z\gamma$ final state is also used to search~\cite{CMS:2022ahq,ATLAS:2026xpl} for the rare Higgs-boson decay $H\to Z\gamma$~\cite{Gehrmann:2015dua,Buccioni:2023qnt}, with the $Z$-boson reconstructed through its leptonic decays.
This loop-induced decay provides a complementary probe of the Higgs couplings to electroweak gauge bosons.
The continuum process $gg\to Z\gamma$ contributes to the background in this search  and can interfere with the Higgs-mediated signal. 
Reliable theoretical predictions for the Standard Model process are therefore needed both for precision comparisons with data and as a baseline for new-physics searches.

The dominant contribution to hadronic $Z \gamma$ production arises from quark-antiquark annihilation. In this channel, the NLO QCD corrections to this process have been studied 
where the $Z$-boson is on-shell~\cite{Ohnemus:1992jn} and for $Z$-bosons decaying leptonically~\cite{DeFlorian:2000sg,Campbell:2011bn}.
The next-to-next-to-leading order (NNLO) QCD corrections have also been computed using  $q_T$-subtraction, where the $Z$-boson is allowed to decay leptonically 
in Refs~\cite{Grazzini:2013bna,Grazzini:2015nwa}.
Here, the first order QCD corrections are estimated to be about $44.3$\%, while the second order corrections are about $9.3$\% for the $13$ TeV ~\cite{Grazzini:2017mhc}. The NLO electroweak correction has also been studied in Ref.~\cite{Denner:2015fca}.
Studies have also been done to include higher-order soft-gluon effects for the quark-initiated channel~\cite{Kidonakis:2026jek}.
At NNLO, the gluon-fusion channel $gg\to Z \gamma$ also contributes to the hadronic cross section via loop-induced processes.
This gluon fusion process has been studied at leading order~\cite{Ametller:1985di,vanderBij:1988fb,Adamson:2002rm}. The large gluon flux at high energy colliders like LHC makes this channel non-negligible even though it is formally suppressed by additional powers of the strong coupling ($\alpha_s$).
This process at the lowest order contributes approximately $1.5$\% of the LO $q\bar{q}$ 
result. The higher
order corrections to the gluon fusion process are generally very large, and hence at higher orders it can contribute even more.
Because of the large gluon fluxes, the 
 higher-order predictions for this loop-induced  process will be important, and they
 formally start contributing at N$^3$LO in the inclusive $Z\gamma$ cross section.
The relative size of its radiative corrections and the associated scale dependence therefore need to be assessed within its own perturbative series.
The NLO calculation provides the first radiative correction for this process. The two-loop helicity amplitudes required for its virtual QCD corrections have been computed in massless QCD~\cite{Gehrmann:2013vga}. However, the full NLO QCD corrections to the loop-induced channel remain unknown.

At higher orders in the perturbation theory, in kinematic regions where the gluon radiation becomes either soft or collinear, the cross sections will be dominated by large contributions arising from the threshold logarithms. To assess the impact of these logarithms, 
one needs to resum them to all orders in perturbation theory.
Threshold enhancements have been studied previously for several quark-antiquark and gluon-initiated processes in soft-virtual in Refs.~\cite{Sterman:1986aj,Catani:1989ne,Catani:1990rp,Catani:1996yz,Contopanagos:1996nh,Catani:2003zt,Eynck:2003fn,Moch:2005ba,Moch:2005ky,Idilbi:2006dg,Ravindran:2005vv,Ravindran:2006cg,Ahmed:2014cla,Ahmed:2014cha,Kumar:2014uwa,Bonvini:2014joa,Catani:2014uta,Ahmed:2016otz,AH:2019phz,Ahmed:2019udm,Das:2019bxi,Das:2019btv,AH:2020cok,Das:2020gie,Ahmed:2020nci,Das:2020pzo,Das:2022zie,Das:2024auk,Banerjee:2024cpr,Das:2025wbj,Banerjee:2025tbo,Das:2025qym,Banerjee:2026hzp,Goyal:2025bzf,Kidonakis:1996aq,Kidonakis:1997gm,Kidonakis:2003tx,Kidonakis:2005kz,Kidonakis:2007ww,Kidonakis:2017dmh,Forslund:2020lnu}  and next-to-soft-virtual correction in Refs.~\cite{AH:2020iki,AH:2021kvg,AH:2021vdc,Bhattacharya:2021hae,AH:2022lpp,Bhattacharya:2025rqk,Pal:2023vec,Pal:2024eyr}.

In this work, we present the full NLO QCD corrections to this loop-induced gluon fusion process. We further resum threshold logarithms to
NLL accuracy and match the resummed prediction to the NLO result. 
We provide the $Z\gamma$ invariant-mass distribution for 
proton-proton collisions at $\sqrt{S}=13.6~\mathrm{TeV}$, together with production cross sections for several collider energies.

This paper is organized as follows. In Section \ref{sec:theory}, we present the theoretical framework required
for the computation of NLO QCD corrections as well as the threshold resummation at NLL accuracy. 
We then present the numerical results in Section \ref{sec:num}, wherein, along with the conventional scale uncertainties, we also present the uncertainties due to the strong coupling constant $\as$ and those
due to the non-perturbative parton distribution functions (PDFs). We also estimate the contribution
of top quark in our calculation using the Born Improved Theory. We then conclude in Section \ref{sec:conclusion}. 

\begin{figure}[t]
  \centering

  \begin{subfigure}[b]{0.32\textwidth}\centering
\begin{tikzpicture}[scale=\diagscale]
  \diagframe
  \coordinate (A) at (0,0.7);  \coordinate (B) at (1.5,0.7);
  \coordinate (C) at (1.5,-0.7); \coordinate (D) at (0,-0.7);
  \draw[gluon] (-2.0,1.5) -- (A);
  \draw[gluon] (-2.0,-1.5) -- (D);
  \draw[zboson] (B) -- (2.8,1.5);
  \draw[photon] (C) -- (2.8,-1.5);
  \draw[fermion] (A)--(B); \draw[fermion] (B)--(C);
  \draw[fermion] (C)--(D); \draw[fermion] (D)--(A);
  \node[lab,left]  at (-2.0,1.5)  {$g$};
  \node[lab,left]  at (-2.0,-1.5) {$g$};
  \node[lab,right] at (2.8,1.5)   {$Z$};
  \node[lab,right] at (2.8,-1.5)  {$\gamma$};
\end{tikzpicture}
    \caption{}\label{fig:diaga}
  \end{subfigure}\hfill
  \begin{subfigure}[b]{0.32\textwidth}\centering
\begin{tikzpicture}[scale=\diagscale]
  \diagframe
  \coordinate (A) at (0,0.7);  \coordinate (B) at (1.6,0.7);
  \coordinate (C) at (1.6,-0.7); \coordinate (D) at (0,-0.7);
  \coordinate (M1) at (0.8,0.7); \coordinate (M2) at (0.8,-0.7);
  \draw[gluon] (-2.0,1.5) -- (A);
  \draw[gluon] (-2.0,-1.5) -- (D);
  \draw[zboson] (B) -- (2.8,1.5);
  \draw[photon] (C) -- (2.8,-1.5);
  \draw[fermion] (A)--(M1); \draw[fermion] (M1)--(B);
  \draw[fermion] (B)--(C);
  \draw[fermion] (C)--(M2); \draw[fermion] (M2)--(D);
  \draw[fermion] (D)--(A);
  \draw[gluon] (M1) -- (M2);
  \node[lab,left]  at (-2.0,1.5)  {$g$};
  \node[lab,left]  at (-2.0,-1.5) {$g$};
  \node[lab,right] at (2.8,1.5)   {$Z$};
  \node[lab,right] at (2.8,-1.5)  {$\gamma$};
\end{tikzpicture}
    \caption{}\label{fig:diagb}
  \end{subfigure}\hfill
  \begin{subfigure}[b]{0.32\textwidth}\centering
\begin{tikzpicture}[scale=\diagscale]
  \diagframe
  \coordinate (P1) at (-0.60,0.14); \coordinate (P2) at (-0.26,-0.92);
  \coordinate (P3) at (0.86,-0.92);  \coordinate (P4) at (1.20,0.14);
  \coordinate (P5) at (0.30,0.80);
  \draw[gluon]  (-1.70,0.50) -- (P1);
  \draw[gluon]  (-0.93,-1.85) -- (P2);
  \draw[photon] (P3) -- (1.53,-1.85);
  \draw[zboson] (P4) -- (2.30,0.50);
  \draw[gluon]  (P5) -- (0.30,1.90);
  \draw[fermion] (P1)--(P5); \draw[fermion] (P5)--(P4);
  \draw[fermion] (P4)--(P3); \draw[fermion] (P3)--(P2);
  \draw[fermion] (P2)--(P1);
  \node[lab,left]  at (-1.70,0.50)  {$g$};
  \node[lab,left]  at (-0.93,-1.85) {$g$};
  \node[lab,right] at (1.53,-1.85)  {$\gamma$};
  \node[lab,right] at (2.30,0.50)   {$Z$};
  \node[lab,above] at (0.30,1.90)   {$g$};
\end{tikzpicture}
    \caption{}\label{fig:diagc}
  \end{subfigure}

  \vspace{3mm}

  \begin{subfigure}[b]{0.32\textwidth}\centering
\begin{tikzpicture}[scale=\diagscale]
  \diagframe
  \coordinate (A) at (0,0.7);  \coordinate (B) at (1.5,0.7);
  \coordinate (C) at (1.5,-0.7); \coordinate (D) at (0,-0.7);
  \coordinate (R) at (-1.3,1.15);
  \draw[gluon] (-2.4,1.5) -- (R);
  \draw[gluon] (R) -- (-0.9,2.0);
  \draw[gluon] (R) -- (A);
  \draw[gluon] (-2.4,-1.5) -- (D);
  \draw[zboson] (B) -- (2.8,1.5);
  \draw[photon] (C) -- (2.8,-1.5);
  \draw[fermion] (A)--(B); \draw[fermion] (B)--(C);
  \draw[fermion] (C)--(D); \draw[fermion] (D)--(A);
  \node[lab,left]  at (-2.4,1.5)  {$g$};
  \node[lab,above] at (-0.9,2.0)  {$g$};
  \node[lab,left]  at (-2.4,-1.5) {$g$};
  \node[lab,right] at (2.8,1.5)   {$Z$};
  \node[lab,right] at (2.8,-1.5)  {$\gamma$};
\end{tikzpicture}
    \caption{}\label{fig:diagd}
  \end{subfigure}\hfill
  \begin{subfigure}[b]{0.32\textwidth}\centering
\begin{tikzpicture}[scale=\diagscale]
  \diagframe
  \coordinate (V) at (0,1.4);
  \coordinate (A) at (0,0.3);  \coordinate (B) at (1.4,0.3);
  \coordinate (C) at (1.4,-1.1); \coordinate (D) at (0,-1.1);
  \draw[fermion] (-2.2,1.9) -- (V);
  \draw[fermion] (V) -- (2.6,1.9);
  \draw[gluon] (V) -- (A);
  \draw[gluon] (-2.2,-1.8) -- (D);
  \draw[zboson] (B) -- (2.8,0.6);
  \draw[photon] (C) -- (2.8,-1.8);
  \draw[fermion] (A)--(B); \draw[fermion] (B)--(C);
  \draw[fermion] (C)--(D); \draw[fermion] (D)--(A);
  \node[lab,left]  at (-2.2,1.9)  {$q$};
  \node[lab,right] at (2.6,1.9)   {$q$};
  \node[lab,left]  at (-2.2,-1.8) {$g$};
  \node[lab,right] at (2.8,0.6)   {$Z$};
  \node[lab,right] at (2.8,-1.8)  {$\gamma$};
\end{tikzpicture}
    \caption{}\label{fig:diage}
  \end{subfigure}\hfill
  \begin{subfigure}[b]{0.32\textwidth}\centering
\begin{tikzpicture}[scale=\diagscale]
  \diagframe
  \coordinate (V1) at (-1.5,1.2); \coordinate (V2) at (0,1.2);
  \coordinate (A) at (0,0.2); \coordinate (B) at (1.3,-0.5);
  \coordinate (D) at (0,-1.2);
  \draw[fermion] (-2.4,1.2) -- (V1);
  \draw[fermion] (V1) -- (V2);
  \draw[fermion] (V2) -- (2.6,1.2);
  \draw[photon] (V1) -- (-0.2,2.35);
  \draw[gluon] (V2) -- (A);
  \draw[gluon] (-2.4,-1.8) -- (D);
  \draw[zboson] (B) -- (2.8,-0.5);
  \draw[fermion] (A)--(B); \draw[fermion] (B)--(D); \draw[fermion] (D)--(A);
  \node[lab,left]  at (-2.4,1.2)  {$q$};
  \node[lab,right] at (2.6,1.2)   {$q$};
  \node[lab,right] at (-0.2,2.35) {$\gamma$};
  \node[lab,left]  at (-2.4,-1.8) {$g$};
  \node[lab,right] at (2.8,-0.5)  {$Z$};
\end{tikzpicture}
    \caption{}\label{fig:diagf}
  \end{subfigure}

  \vspace{3mm}

  \begin{subfigure}[b]{0.32\textwidth}\centering
\begin{tikzpicture}[scale=\diagscale]
  \diagframe
  \coordinate (V1) at (0,1.5); \coordinate (V2) at (0.9,1.55);
  \coordinate (A) at (0,0.2); \coordinate (B) at (1.3,-0.5);
  \coordinate (D) at (0,-1.2);
  \draw[fermion] (-2.4,1.5) -- (V1);
  \draw[fermion] (V1) -- (V2);
  \draw[fermion] (V2) -- (2.8,0.95);
  \draw[photon] (V2) -- (2.8,2.25);
  \draw[gluon] (V1) -- (A);
  \draw[gluon] (-2.4,-1.8) -- (D);
  \draw[zboson] (B) -- (2.8,-0.5);
  \draw[fermion] (A)--(B); \draw[fermion] (B)--(D); \draw[fermion] (D)--(A);
  \node[lab,left]  at (-2.4,1.5)  {$q$};
  \node[lab,right] at (2.8,0.95)  {$q$};
  \node[lab,right] at (2.8,2.25)  {$\gamma$};
  \node[lab,left]  at (-2.4,-1.8) {$g$};
  \node[lab,right] at (2.8,-0.5)  {$Z$};
\end{tikzpicture}
    \caption{}\label{fig:diagg}
  \end{subfigure}\hfill
  \begin{subfigure}[b]{0.32\textwidth}\centering
\begin{tikzpicture}[scale=\diagscale]
  \diagframe
  \coordinate (V) at (-0.8,0); \coordinate (G) at (0.4,0);
  \coordinate (T) at (1.5,0.8); \coordinate (Zv) at (1.5,0);
  \coordinate (Bv) at (1.5,-0.8);
  \draw[fermion] (-2.4,-1.4) -- (V);
  \draw[fermion] (V) -- (-2.4,1.4);
  \draw[gluon] (V) -- (G);
  \draw[fermion] (G)--(Bv); \draw[fermion] (Bv)--(Zv);
  \draw[fermion] (Zv)--(T); \draw[fermion] (T)--(G);
  \draw[gluon]  (T)  -- (2.6,1.4);
  \draw[zboson] (Zv) -- (2.8,0);
  \draw[photon] (Bv) -- (2.6,-1.4);
  \node[lab,left]  at (-2.4,-1.4) {$q$};
  \node[lab,left]  at (-2.4,1.4)  {$\bar q$};
  \node[lab,right] at (2.6,1.4)   {$g$};
  \node[lab,right] at (2.8,0)     {$Z$};
  \node[lab,right] at (2.6,-1.4)  {$\gamma$};
\end{tikzpicture}
    \caption{}\label{fig:diagh}
  \end{subfigure}\hfill
  \begin{subfigure}[b]{0.32\textwidth}\centering
\begin{tikzpicture}[scale=\diagscale]
  \diagframe
  \coordinate (P) at (-0.9,-1.0); \coordinate (V) at (-0.9,0.5);
  \coordinate (G) at (0.5,0.5);
  \coordinate (Zv) at (1.7,1.0); \coordinate (Gv) at (1.7,0.0);
  \draw[fermion] (-2.4,-1.7) -- (P);
  \draw[fermion] (P) -- (V);
  \draw[fermion] (V) -- (-2.4,1.6);
  \draw[photon] (P) -- (0.6,-1.8);
  \draw[gluon]  (V) -- (G);
  \draw[fermion] (G)--(Gv); \draw[fermion] (Gv)--(Zv); \draw[fermion] (Zv)--(G);
  \draw[zboson] (Zv) -- (2.8,1.4);
  \draw[gluon]  (Gv) -- (2.8,0.0);
  \node[lab,left]  at (-2.4,-1.7) {$q$};
  \node[lab,left]  at (-2.4,1.6)  {$\bar q$};
  \node[lab,right] at (0.6,-1.8)  {$\gamma$};
  \node[lab,right] at (2.8,1.4)   {$Z$};
  \node[lab,right] at (2.8,0.0)   {$g$};
\end{tikzpicture}
    \caption{}\label{fig:diagi}
  \end{subfigure}

  \vspace{2mm}
  \caption{Representative diagrams for $gg\to Z\gamma$ at NLO: (a) LO quark box; (b) two-loop virtual; (c),(d) real gluon emission in the $gg$ channel; (e)--(g) $qg$ channel; (h),(i) $q\bar q$ channel.}
  \label{fig:feyn}
\end{figure}
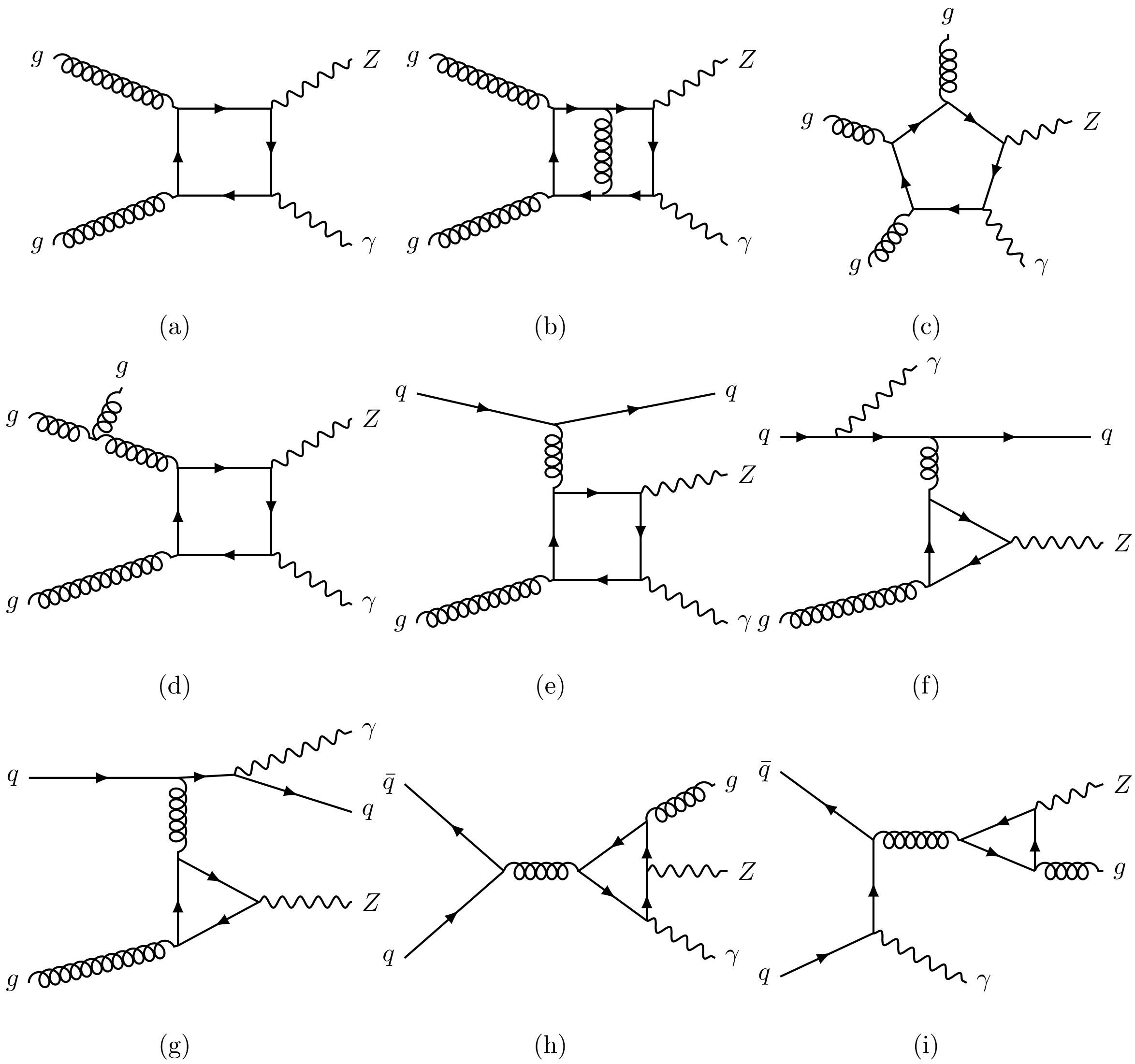

\section{Theoretical Framework}\label{sec:theory}

We consider the loop-induced contribution to $Z\gamma$ production.
The Born-level process is
\begin{equation}
  g(p_1)+g(p_2)\longrightarrow Z(p_3)+\gamma(p_4),
  \label{eq:zgamma-process}
\end{equation}
with $p_1^2=p_2^2=p_4^2=0$ and $p_3^2=m_Z^2$.
The hadronic cross section for producing a $Z\gamma$ pair in proton--proton collisions is given by the QCD factorization formula,
\begin{align}\label{eq:had-xsect}
 \frac{\df \sigma}{\df Q}  
=  
\sum_{a,b}\int_0^1 \df x_1\int_0^1 \df x_2 \,\,f_{a}(x_1,\mu_F^2)\,
f_{b}(x_2,\mu_F^2) 
\int_0^1 \df z~ \delta \left(\tau-zx_1 x_2\right)
\frac{\df \hat\sigma_{ab}}{\df Q}
\, .
\end{align}
Here, $f_a$ and $f_b$ denote the PDFs of the partons $a$ and $b$ in the incoming protons, and $\hat{\sigma}_{ab}$ is the parton-level cross-section. The dimensionless variables $z=Q^2/s$ and $\tau=Q^2/S$ describe the partonic and hadronic thresholds, respectively. The quantities $\sqrt{S}$ and $\sqrt{s}$ represent the hadronic and partonic center-of-mass energies respectively, while $Q$ is the invariant mass of the $Z\gamma$ pair and $\muf$ is the factorization scale.

\subsection{Fixed-order calculation}

 The partonic cross section at NLO in QCD can be written as
\begin{align}\label{eq:PARTONIC-COEFF}
\df \hat\sigma_{\rm NLO}
&=
\df \hat\sigma^{(0)}
+ \df \hat\sigma_{\rm virt}
+ \df \hat\sigma_{\rm real}
\, .
\end{align}

It proceeds only through a closed quark loop, so the Born ($\hat\sigma^{(0)}$) contribution already starts at $\mathcal{O}(\as^2)$.
At NLO, i.e.\ at $\mathcal{O}(\as^3)$, the virtual corrections
$\df \hat\sigma_{\rm virt}$ come from the interference of this one-loop Born amplitude with the two-loop  
amplitude, see Figure~\ref{fig:diaga} and Figure~\ref{fig:diagb} for representative diagrams.
We take the latter with massless internal quarks from Ref.~\cite{Gehrmann:2013vga}, where helicity amplitudes are presented for the decay-channel process ($gg\to\gamma V^*\to\gamma\ell^+\ell^-$), with the production current contracted with the leptonic decay current. To obtain the corresponding on-shell $Z\gamma$ contribution, we set $q^2 = m_Z^2$, remove the $Z$-boson propagator and the appropriate ($Z\ell\ell$) coupling factors, and use the resulting decay-channel helicity amplitudes to implement the complete on-shell $Z$-polarization sum.
This conversion follows from the massive-vector polarization completeness relation discussed in Refs.~\cite{Dixon:1998py,Peskin:2017emn,DeLaurentis:2026dpj}. 
Following this procedure, we obtain loop amplitudes for the case where $Z$-boson is on-shell from the results of Ref.~\cite{Gehrmann:2013vga}. 
As a validation, the one-loop squared matrix element was compared numerically with an independent result obtained from { \tt MadGraph5\_aMC@NLO}~\cite{Alwall_2011} at several phase-space points, keeping the $Z$ boson on-shell.
The numerical agreement was found up to $14$ decimal places, with residual differences consistent with the numerical precision of MadLoop~\cite{Hirschi:2011pa}.

The real corrections $\df \hat\sigma_{\rm real}$ are obtained by squaring one-loop
$2 \to 3$ amplitudes. They receive contributions from three different subprocesses: $gg \to Z\gamma g$
(Figures~\ref{fig:diagc} and~\ref{fig:diagd}), $qg \to Z\gamma q$ and $gq \to Z\gamma q$
(Figures~\ref{fig:diage}--\ref{fig:diagf}) and $q\bar q \to Z\gamma g$
(Figures~\ref{fig:diagh} and~\ref{fig:diagi}). All Born and real-emission matrix
elements are evaluated numerically 
using \texttt{RECOLA}~\cite{Actis:2016mpe}.

The phase space integration of the real emission diagrams gives rise to  infrared (IR) divergences. 
In the $gg$ initiated subprocess,  they appear when the final-state gluon becomes soft, or when it becomes
collinear to one of the incoming gluons. In the $qg$ subprocess, the divergence is purely
collinear: it arises when the final-state quark is collinear to the incoming quark.
On the contrary, the $q\bar{q}$ initiated subprocess will have only finite contribution without any IR divergences.
We cancel all of these divergences using the Catani--Seymour dipole subtraction
method~\cite{Catani:1996vz}, implemented in an in-house code.
We have validated this implementation against the \texttt{ggxy}~\cite{Davies:2026uxl}
package for $ZH$ production~\cite{Banerjee:2026hzp}.

\subsection{Threshold Resummation}

The partonic cross section in ~\eq{eq:had-xsect} can be written in terms of the partonic coefficient $\Delta_{ab}(z)$ as follows:
\begin{align}\label{eq:PARTONIC-DECOMPOSE1}
\frac{\df \hat\sigma_{ab}}{\df Q}
& =
\hat{\sigma}^{(0)}(Q^2)  
\Delta_{ab} (z)\,.
\end{align} 
Near the partonic threshold, most of the available energy goes
into the $Z\gamma$ pair, leaving little room for additional QCD radiation so that the gluons become essentially soft.
In general, one can write the partonic coefficient  as the sum of soft-virtual (SV) and regular (REG) terms (for brevity, we omit the scales in the
arguments), given as:
\begin{align}\label{eq:PARTONIC-DECOMPOSE2}
\frac{\df \hat\sigma_{ab}(z)}{\df Q}
& =
\hat{\sigma}^{(0)}_{ab}(Q^2) \Big( 
\delta_{b {a}}\Delta_{ab}^{\rm SV}\left(z\right) 
+ \Delta_{ab}^{\rm REG}\left(z\right)
\Big) \,.
\end{align}
The soft-virtual part, \(\Delta_{ab}^{\rm SV}\), collects the
singular contributions near \(z\to1\). They come from unresolved
radiation and virtual corrections to the Born process. 
The regular part, \(\Delta_{ab}^{\rm REG}\), includes
terms that result from the
hard non-collinear parton radiation.
It contains logarithms of $(1-z)$ as well as \((1-z)^n\) type terms, which remain integrable at threshold.
Both parts are expanded in the strong coupling
\(\alpha_s(\mur^2) \equiv 4 \pi a_s(\mur^2)\), with \(\mur\) being the
renormalization scale:
\begin{align}\label{eq:SVREG-Expansion}
\Delta^{\rm SV}_{ab}(z) 
&= \sum_{n=0}^{\infty}\as^n(\mur^2) ~\delta_{ab}
\left( 
\Delta^{(n)}_{\delta} ~\delta(1-z) + 
\sum_{k=0}^{2n-1}  \Delta^{(n)}_{{\cal D}_k} ~{\cal D}_k(1-z) 
\right)\,, ~ 
\text{with } ab \in \{gg, q\bar{q} \} \,,
\nonumber
\\
\Delta^{\rm REG}_{ab}(z) 
&= 
\sum_{n=1}^{\infty}\as^n(\mur^2)
\Delta_{ab}^{(n)}(z)\,,

\end{align}

Here \(\delta(1-z)\) is the Dirac delta distribution, and
\({\cal D}_k(1-z) \equiv \left[\ln^k(1-z)/(1-z)\right]_+\) is a
plus distribution.
The SV coefficient combines the virtual correction to the hard
amplitude, the soft function~\cite{Ravindran:2005vv,
Ravindran:2006cg,
Sudakov:1954sw,
Mueller:1979ih,
Collins:1980ih,
Sen:1981sd} for unresolved radiation, and the
mass-factorization kernels
\cite{Moch:2004pa,Vogt:2004mw}. The virtual correction
and the soft function contain infrared poles. These poles cancel
in the SV coefficient after mass factorization.

We resum the threshold terms by taking the Mellin transform of the cross section from $z$ to $N$ space, where the threshold limit \(z\to1\) now corresponds to \(N\to\infty\).
In Mellin space, the plus distributions give powers of $\ln N$ terms, the terms of the order  $\mathcal{O}(1/N)$
and constant terms. The logarithms are resummed in the large-$N$
limit, while terms suppressed by powers of $1/N$ are neglected
in the SV approximation. In the Mellin space, the SV coefficient can be written as
\begin{align}\label{eq:resum-partonic}
	\frac{\df \hat{\sigma}_{N,ab}^{\rm N^{\it n}LL}}{\df Q}
	= \int_0^1 \df z ~ z^{N-1} \left(  \Delta^{\rm SV}_{ab}(z) \right)
	\equiv g_{0}(Q^2) \exp \left( G_N \right)\,.
    
\end{align}
The prefactor \(g_0\) contains terms independent of \(N\) whereas the exponent \(G_N\) collects the large threshold logarithms. 
The expansion of $G_N$ reads as follows~\cite{Moch:2005ba,Catani:2003zt}
\begin{align}\label{eq:GN}
G_N &= 
	\ln (\Nbar) ~g_1(\omega)
        + \sum_{n=1}^{\infty} \as^{n-1}(\mur^2)~ g_{n+1}(\omega) \,,
\end{align}
Here \(\Nbar = N \exp (\gamma_E)\), with \(\gamma_E\) the Euler--Mascheroni constant, and \(\omega = 2 \beta_0 \as(\mur^2) \ln \Nbar\).
The function \(g_1\) gives the leading logarithms (LL), while
\(g_2\) adds the next-to-leading logarithms (NLL). 
Further terms give higher logarithmic accuracy successively. 
These terms are universal and depend on the type of incoming partons.

For the NLL prediction, we keep $G_N$ up to the \(g_2\) term in expansion~\eq{eq:GN}. These are well known in the literature~\cite{Catani:2003zt,Moch:2005ba}. 
For the present gluon initiated case we have:
\begin{align}
    g_1(\omega) &= 
    \frac{1}{\beta_0} \bigg\{ 
                             C_A \left( 8 \left( 1+  \frac{\wb}{\omega}\Lw \right) \right)
                    \bigg\}
    \,,
    \nn
    g_2(\omega) &= \frac{1}{\beta_0^3} \bigg\{ 
            C_A \beta_1 \left( 4 \omega + 4 \Lw + 2 \Lw^2\right)
            + C_A n_f \beta_0 \left(\frac{40}{9} \left( \omega + \Lw\right) \right)
            \nn
            &\hspace{2.5em}+ C_A^2 \beta_0 \left( 
                                    \left(-\frac{268}{9} + 8 \zeta_2 \right) \left(\omega + \Lw\right) 
                            \right) 
            + C_A \beta_0^2 \left( 4 \Lw \ln  \left(\frac{Q^2}{\mur^2}\right) + 4 \omega \ln\left( \frac{\muf^2}{\mur^2}\right) 
 \right)
    \bigg\},
\end{align}
In these expressions, \(\wb = 1-\omega\) and
\(\Lw = \ln(\wb)\). 
The process dependent coefficient $g_0$ is also expanded in the strong coupling as,
\begin{align}
	g_0(Q^2) = 1+\sum_{n=1}^{\infty} \as^n(\mur^2) ~g_{_{0n}}(Q^2) \,.
\end{align}
Here we keep $g_{0}$ up to the $\mathcal{O}(\as)$, the $g_{01}$ for any $2 \to n$ color-singlet process can be written as ~\cite{Ahmed:2020nci} ,
\begin{align}\label{eq:g01}
{g}_{_{01}}(Q^2) &=  2\mathcal{M}_{(0,1)} 
        - 2 \beta_{0} \ln\left(\frac{\muf^2}{\mur^2}\right) 
        + 2 \Ca \ln^{2} \left(\frac{Q^2}{\mur^2}\right) 
        + 2 \Ca \zeta_{2}
\end{align}
The $g_{01}$ gets its process dependence from the two-loop IR-finite amplitude ($\mathcal{M}_{(0,1)}$) \cite{Gehrmann:2013vga}.
The resummed result in ~\eq{eq:resum-partonic} is matched with the fixed-order result to account for the hard regular terms and avoid double counting of the SV logarithms. 
While performing the Mellin inversion, one encounters Landau pole. To avoid this problem, one chooses the contour using the minimal prescription
\cite{Catani:1996yz}.
The integration contour is parametrized as $N = c + x e^{i\phi}$, where \(x\) is a real variable. The parameter \(c\) is chosen such that all singularities, except the Landau pole, lie to the left of the contour. In our numerical analysis, we take \(c=1.9\) and \(\phi=3\pi/4\).
The matched result is,
\begin{align}\label{eq:MATCHING}
	\frac{\df \sigma^{\rm N^{\it n}LO+N^{\it n}LL}_{ab}}{\df Q}
	=&
	\frac{\df {\sigma}^{\rm N^{\it n}LO}_{ab} }{\df Q}
	+
        \sum_{ab \in \{gg, q\bar{q}\} }
        \hat{\sigma}^{(0)}_{ab}(Q^2)
	\int_{c-i\infty}^{c+i\infty}
	\frac{\df N}{2\pi i}
	\tau^{-N}
	f_{a,N}(\muf^2)
	f_{b,N}(\muf^2)
	\nn
	&\times 
	\Bigg( 
		\frac{\df \hat{\sigma}_{N,ab}^{\rm N^{\it n}LL}}{\df Q} 
		- 	
		\frac{\df \hat{\sigma}_{N,ab}^{\rm N^{\it n}LL}}{\df Q} \Bigg|_{\rm tr}
	\Bigg) \,.

\end{align}

In ~\eq{eq:MATCHING}, \(f_{a,N}\) and \(f_{b,N}\) are the $N$-space parton distributions.  
They can be evolved with \textsc{QCD-PEGASUS}~\cite{Vogt:2004ns} or obtained from PDFs in $z$-space 
\cite{Kulesza:2002rh}.
The subscript \(\it{tr}\) denotes the expansion of the resummed coefficient through the order of the fixed-order calculation.  
At NLO$+$NLL, the subtraction removes the SV terms already included at NLO.
The matched result thus retains the full NLO contribution and adds higher-order threshold logarithms in the gluon channel.
We next study the effect
of this threshold resummation on the production cross section as well as the invariant mass distribution of $Z\gamma$ 
at the LHC.

\section{Numerical Results}\label{sec:num}
In this section, we present numerical predictions for $Z\gamma$ production via the loop-induced gluon-fusion channel at hadron colliders up to NLO in QCD, matched with soft-gluon threshold resummation at NLL accuracy. We begin by detailing the numerical setup and input parameters. Subsequently, we present the total hadronic cross sections at the 13.6~TeV LHC along with the total production cross-section predictions from 7~TeV to 14~TeV. Finally, we discuss the invariant-mass differential distributions and evaluate the theoretical uncertainties arising from scale variations, parton distribution functions (PDFs), and the strong coupling constant $\alpha_s$.

For our numerical implementation, we take the physical masses of the electroweak gauge bosons to be $m_Z = 91.188~\mathrm{GeV}$ and  $m_W = 80.3692~\mathrm{GeV}.$
The electroweak coupling in the $G_\mu$ scheme is taken to be
$G_\mu = 1.2043993808 \times 10^{-5}~\mathrm{GeV}^{-2}$ with the corresponding
$\alpha^{-1} = 127.93$.
Unless mentioned otherwise, we consider only the massless quarks in the loops with $n_f=5$.

This choice matches the analytic  five-flavor massless QCD framework of the two-loop virtual helicity amplitudes from Ref.~\cite{Gehrmann:2013vga}.
To ensure infrared safety and to avoid non-perturbative collinear photon fragmentation, we apply the Frixione smooth-cone isolation criterion~\cite{Frixione:1998jh} to the photon. Inside a cone of angular radius $\delta \le \delta_0$ centered on the photon in the $(\eta, \phi)$ plane, the total hadronic transverse energy $E_T(\delta)$ must satisfy
\begin{equation}
E_T(\delta) \le \epsilon ~ p_{T}^{\gamma} \left( \frac{1 - \cos\delta}{1 - \cos\delta_0} \right)^n,
\end{equation}
with cone radius $\delta_0 = 0.4$, exponent $n = 1$, efficiency parameter $\epsilon = 0.5$ and maximum allowed transverse energy $E_{T,\max} = 5$~GeV. In addition, we use transverse-momentum cut for the photon to be $p_T^\gamma > 20~\mathrm{GeV}$ while no explicit kinematic cuts are imposed on the on-shell $Z$ boson.

For our numerical predictions, we use
\texttt{PDF4LHC21\_40\_pdfas}~\cite{PDF4LHCWorkingGroup:2022cjn} 
 taken from the \texttt{LHAPDF},
 with the corresponding
 $\alpha_s(m_Z) = 0.118$.
 The unphysical  renormalization ($\mu_{R}$) and factorization ($\mu_{F}$) scales are set for the central choice to be equal to the invariant mass $Q$ of the $Z\gamma$ system, 
 $\mu_{R} =\mu_{F} = Q$.
To determine the theoretical uncertainties due to missing higher-order QCD corrections, we perform the standard seven-point scale variation by varying $\mu_R$ and $\mu_F$ independently by a factor of two around the central scale $Q$ subject to $1/2 \le \mu_R/\mu_F \le 2$.
The scale uncertainty is quantified as
(for $i=$ LO, LO+LL, NLO and NLO+NLL),
\begin{align}
	\mu = \frac{\sigma_{i}(\mur,\muf) - \sigma_{i}(Q,Q)}{\sigma_{i}(Q,Q)} \times 100 \%
    \label{eq:scale_unc}
\end{align}
The PDF uncertainties are computed using the 40 Hessian members of the \\ $\texttt{PDF4LHC21\_40\_pdfas}$ set, following the standard prescription implemented in the 
$\tt{LHAPDF6}$ \cite{Buckley:2014ana}. 
For uncertainty arising from $\alpha_{s}$ we follow the PDF4LHC recommendations given in\cite{PDF4LHCWorkingGroup:2022cjn}.
A 1$\sigma$ variation of $\pm0.0015$ is applied to estimate the associated uncertainty, computed as
\begin{equation}
\delta(\alpha_{s})=\pm3/4|\sigma(\alpha_{s}^{+}(m_{Z}))-\sigma(\alpha_{s}^{-}(m_{Z}))|/\sigma(\alpha_{s}^{c}(m_{Z})),
\end{equation}
where 
$\alpha_{s}^{-}(m_{Z})=0.117$, $\alpha_{s}^{+}(m_{Z})=0.119$ and
$\alpha_{s}^{c}(m_{Z})=0.118$.
We combine the uncertainties from PDF and $\alpha_{s}$ by adding the respective errors in quadrature. In Table~\ref{tab:tableggZA_pdfuncertainty}, we present the total production cross sections 
at the LHC center-of-mass energy of $\sqrt{S} = 13.6$~TeV at LO, NLO, and matched NLO+NLL accuracy, along with their respective scale, PDF, and $\alpha_s$ uncertainties.
\begin{table}[h]
\centering
\renewcommand{\arraystretch}{1.3}
\begin{tabular}{|c|c|c|c|c|c|}
\hline
Order & $\sigma$ (fb) & Scale (\%) & $\delta(\mathrm{PDF})$ (\%) & $\delta(\alpha_s)$ (\%) & $\delta(\mathrm{PDF}\oplus\alpha_s)$ (\%) \\
\hline
\hline
$\sigma^{\mathrm{LO}}$      & 841.36  & $[+22.3,\,-16.7]$ & $\pm 1.58$ & $\pm 1.33$ & $\pm 2.07$ \\
\hline
$\sigma^{\mathrm{NLO}}$     & 1537.91 & $[+15.2,\,-12.1]$ & $\pm 1.63$ & $\pm 1.74$ & $\pm 2.38$ \\
\hline
$\sigma^{\mathrm{NLO+NLL}}$ & 1887.01 & $[+26.9,\,-19.3]$ & $\pm 1.62$ & $\pm 2.14$ & $\pm 2.68$ \\
\hline
\end{tabular}
\caption{\small{The $gg\to Z\gamma$ production cross sections at LO, NLO and NLO+NLL for $\sqrt{S} = 13.6$ TeV. Uncertainties due to seven-point scale variations, PDFs and $\alpha_s$ are also shown.}}
\label{tab:tableggZA_pdfuncertainty}
\end{table}

At leading order ($\mathcal{O}(\alpha_s^2)$), the cross section is $841.36$~fb. The NLO QCD corrections substantially enhance the cross section by about $83$\%, with  an inclusive NLO $K$-factor of $K_{\text{NLO}} = \sigma^{\text{NLO}}/\sigma^{\text{LO}} \approx 1.83$.

Such large corrections are a well-known behaviour of loop-induced gluon-initiated processes, driven by the color factor $C_A$ in the gluon channels and the opening of new initial-state quark-gluon channels at NLO. However, the seven-point scale variation are found to get reduced from $22.3\%$ at LO to $15.2\%$ at NLO, indicating improved perturbative stability.

Threshold resummation at NLL 
further enhances the total cross section approximately by an additional 22.7\% over the fixed-order NLO prediction, and yielding an overall $K$-factor of $R_{\text{10}} = \sigma^{\text{NLO+NLL}}/\sigma^{\text{LO}} \approx 2.24$. 
The scale uncertainty at NLO+NLL is 
$26.9\%$. The broadening of the scale uncertainty when matching NLL resummation to NLO is known \cite{Banerjee:2026hzp,Banerjee:2024cpr,Banerjee:2025tbo} 
for the inclusive predictions as the total cross section is dominated by the lower-$Q$ region where the resummed results  have the
factorization-scale dependence
up to NLO only while the dependence on $\mu_{R}$ is considered to all orders
in the perturbation theory.

As shown in Table~\ref{tab:tableggZA_pdfuncertainty}, the non-perturbative PDF uncertainties are very small, about $1.6$\%, and are roughly found to be the same across different perturbative orders considered here.
On the other hand, the parametric $\alpha_s$ uncertainty is estimated to be small and is found to vary from $\pm 1.33\%$ at LO
to $\pm 2.14\%$ at NLO+NLL, leading to a combined $\delta(\text{PDF}+\alpha_s)$ uncertainty between $\pm 2.07\%$ and $\pm 2.68\%$. Comparing these with the scale variation shows that the theoretical uncertainty is mostly dominated by the perturbative scale choice, with PDF and $\alpha_s$ uncertainties playing a subdominant role.

To explore the collider energy dependence, we display in Table~\ref{tab:escan} the cross sections at several hadron collider energies, namely $\sqrt{S} = 7, 8, 13$, and $14$~TeV, from 
LO to NLO+NLL.

\setlength{\tabcolsep}{10pt}       
\renewcommand{\arraystretch}{1.5}   
\begin{table}[htbp]
\begin{tabular}{|c|c|c|c|c|}
\hline
$\sqrt{S}$ [TeV] & LO & LO+LL & NLO & NLO+NLL \\
\hline
\hline
7 & $241.04^{+24.5\%}_{-18.8\%}$ & $343.79^{+27.9\%}_{-20.6\%}$ & $461.25^{+16.5\%}_{-12.9\%}$ & $580.14^{+25.5\%}_{-18.1\%}$ \\
\hline
8 & $314.84^{+23.2\%}_{-18.1\%}$ & $444.70^{+27.2\%}_{-19.9\%}$ & $596.94^{+16.0\%}_{-12.6\%}$ & $746.37^{+25.6\%}_{-18.4\%}$ \\
\hline
13 & $777.28^{+22.3\%}_{-16.7\%}$ & $1065.80^{+26.7\%}_{-19.0\%}$ & $1429.21^{+15.2\%}_{-12.2\%}$ & $1755.13^{+26.7\%}_{-19.3\%}$ \\
\hline
14 & $884.95^{+22.3\%}_{-16.7\%}$ & $1208.64^{+26.6\%}_{-19.0\%}$ & $1622.05^{+15.5\%}_{-12.0\%}$ & $1986.91^{+27.1\%}_{-19.1\%}$ \\
\hline
\end{tabular}
\caption{ Total production cross-sections (in fb) at LO, LO+LL, NLO and NLO+NLL for several collider energies, along with their corresponding scale uncertainties relative to the central scale choice.}
\label{tab:escan}
\end{table}

As the collision energy increases from $\sqrt{S} = 7$~TeV to $14$~TeV, the cross section increases by a factor of nearly $3.7 (3.4)$ at LO (NLO+NLL). 
This increase in the cross section is due to the rapid increase in the gluon flux at smaller values of the partonic momentum fractions $x_1$ and $x_2$.
Over this energy range, the NLO corrections enhance the LO cross section by about $83\text{--}91\%$, and threshold resummation consistently contributes an additional $22\text{--}26\%$ enhancement over NLO 
\cite{Das:2025qym,Das:2025wbj, Banerjee:2026hzp}.

\begin{figure}[!htbp]
\centering
\includegraphics[width=\textwidth]{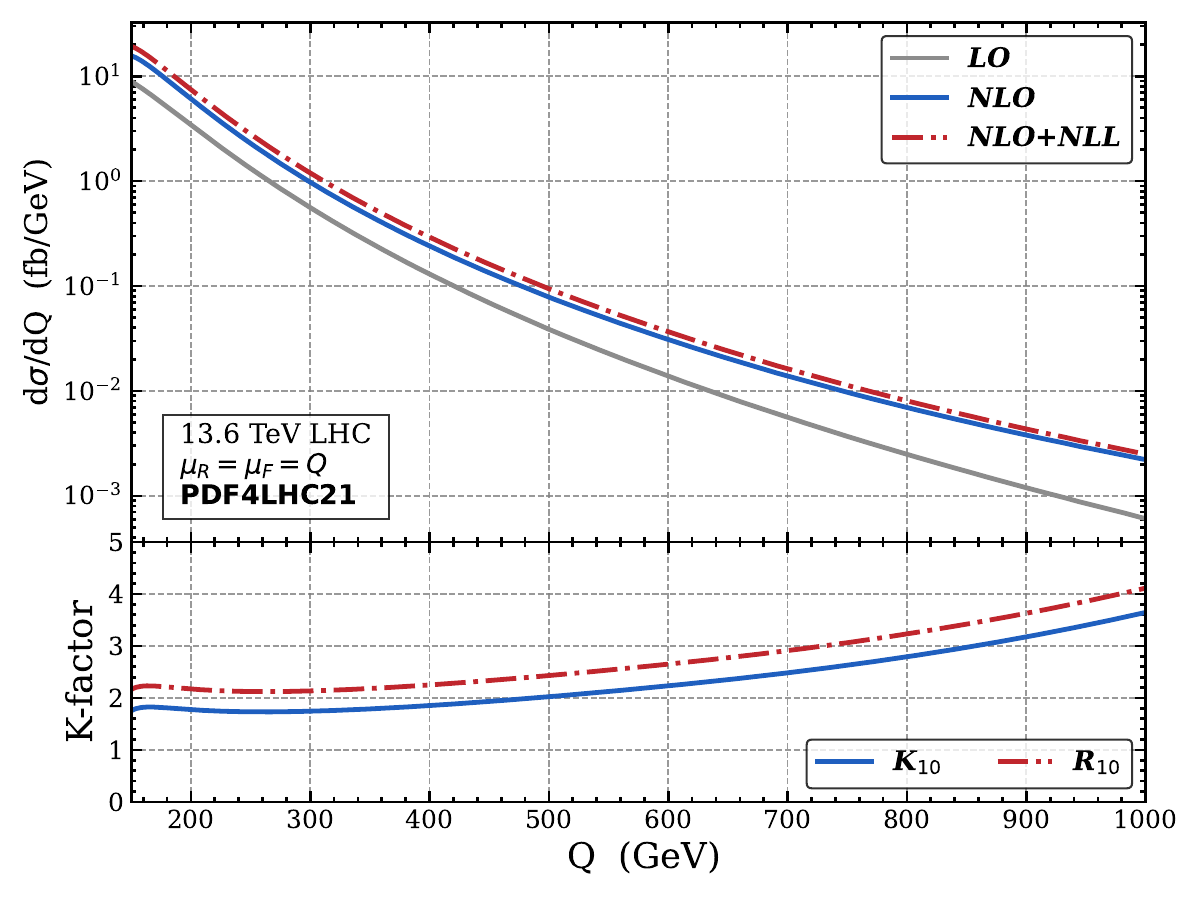}
\caption{Invariant-mass distribution of the $Z\gamma$ system for the $13.6$~TeV LHC at LO, NLO, and NLO+NLL (upper panel), along with the corresponding
K-factors defined in \eq{eq:kfactorQ}
(lower panel).}
\label{fig:invariant-mass-distribution}
\end{figure}

In the rest of the analysis, we present our results for $\sqrt{S} = 13.6$ TeV. We now present the invariant-mass distribution $d\sigma/dQ$ of the $Z\gamma$ system for $Q \in [150, 1000\text{ GeV}]$. The results are displayed in Figure~\ref{fig:invariant-mass-distribution}.
As observed in the upper panel, the differential cross section falls monotonically by more than four orders of magnitude over the range of $Q$ considered, as the parton luminosity drops sharply with increasing $\tau$ .
The lower panel illustrates the $K$-factors for the distribution, defined as
\begin{equation}
K_{10}(Q) = \frac{\df\sigma^{\mathrm{NLO}}/\df Q}{\df\sigma^{\mathrm{LO}}/\df Q}\,, \qquad R_{10}(Q) = \frac{\df\sigma^{\mathrm{NLO+NLL}}/\df Q}{\df\sigma^{\mathrm{LO}}/\df Q}\,.
\label{eq:kfactorQ}
\end{equation}
In the low $Q$ region ($Q \sim 150\text{--}250$~GeV), the fixed-order NLO $K$-factor is around $K_{10} \approx 1.7\text{--}1.8$, and the resummed ratio is $R_{10} \approx 2.1\text{--}2.2$. As $Q$ increases, both 
the fixed order and resummed K-factors grow steeply: $K_{10}$ increases to $\approx 3.6$, while $R_{10}$ reaches $\approx 4.1$.

This rapid growth of K-factors in the high invariant-mass region is precisely due to the channels like $qg$ in the fixed order NLO calculation, where photon is emitted from the quark line in the final state. In the limit, where photon is collinear to the quark, it gives rise to large contributions and eventually to QED collinear singularity. 
However, the QED collinear configuration is removed by the smooth cone isolation algorithm described before.
It is also worth noting that the rising K-factors are due to the relatively large $qg$ flux compared to the $gg$ flux in the high invariant mass region.

\begin{figure}[!htbp]
\centering
\includegraphics[width=\textwidth]{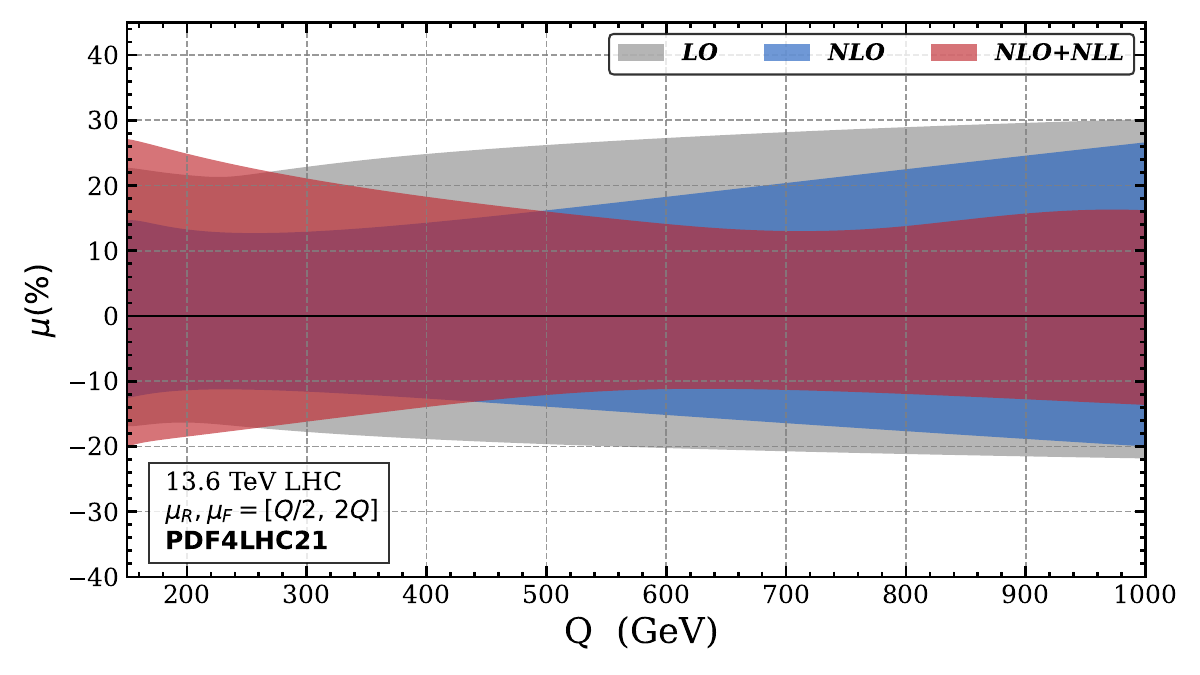}
\caption{Seven-point scale uncertainty bands, defined in 
\eq{eq:scale_unc}, for the $Z\gamma$ production cross section at the $13.6$~TeV LHC as a function of the invariant mass $Q$.}
\label{fig:scale-uncertainty-band}
\end{figure}

To evaluate the impact of higher orders and resummation on perturbative stability, we show in Figure~\ref{fig:scale-uncertainty-band} the $7$-point scale-uncertainties in the invariant mass distribution for LO, NLO, and NLO+NLL normalized to the respective prediction obtained with the choice of central scale.
In the lower invariant-mass regime ($Q \lesssim 300$~GeV), the NLO scale uncertainty band is noticeably narrower than the LO band, shrinking the variation from $[-17\%, +23\%]$ at LO down to $[-15\%, +13\%]$ at NLO. However, as $Q$ increases towards $1000$~GeV, the fixed-order NLO scale uncertainty band broadens, expanding to $[-15\%, +25\%]$ at $Q = 1000$~GeV.
This increase in scale uncertainty at large $Q$ is caused by large contributions from threshold logarithms, which endanger the convergence of the fixed-order expansion.
\begin{figure}[!htbp]
\includegraphics[width=\textwidth]{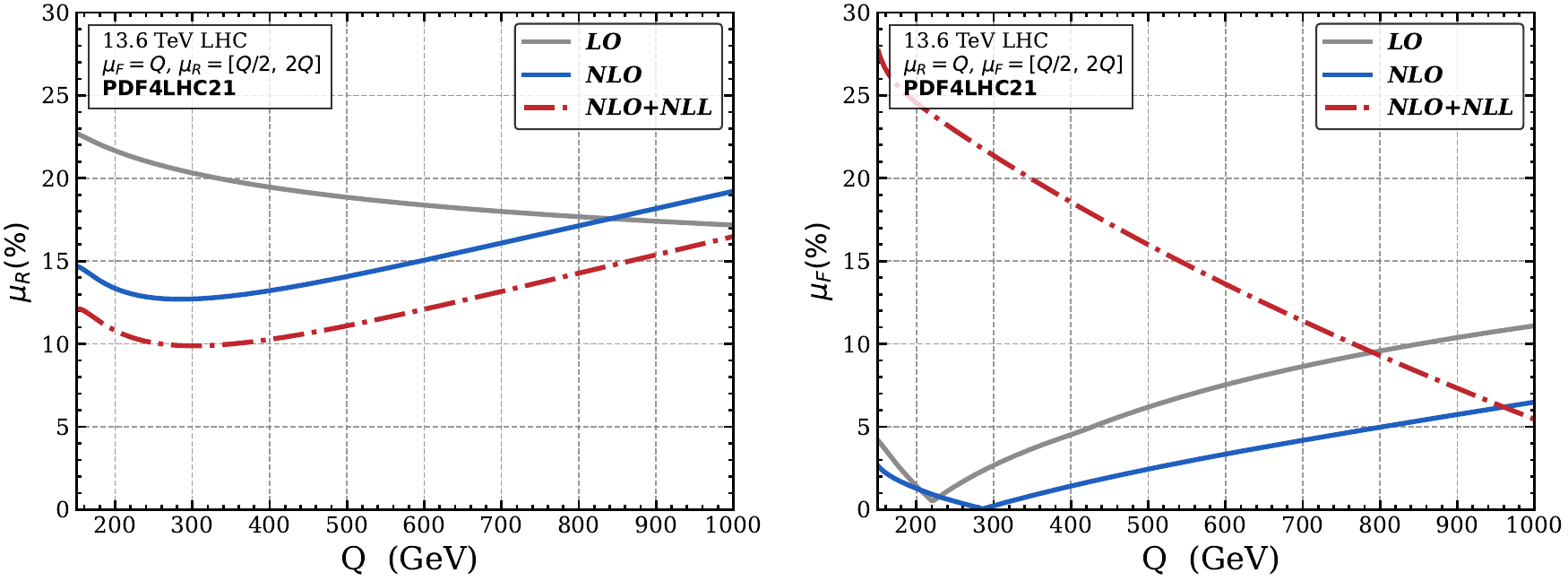}
\caption{The percentage uncertainty due to renormalization (left) and factorization scale (right) as a function of the invariant mass $Q$ at different orders. }
\label{fig:scale-uncertainty-mur-muf-max}
\end{figure}

In contrast, the resummed NLO+NLL predictions are more stable against the scale variations for $Q \gtrsim 500$~GeV. The NLO+NLL scale-uncertainty band contracts significantly compared to NLO, covering only $[-14\%, +16\%]$ at $Q = 1000$~GeV. 
This illustrates the importance of soft-gluon resummation in stabilizing theoretical predictions in the high-$Q$ region.

To precisely see the source of scale uncertainty we represent in Figure~\ref{fig:scale-uncertainty-mur-muf-max} the variation by keeping one of the scales fixed at its central choice and changing the other between $Q/2$ and $2Q$. 
We note that, due to threshold resummation, the renormalization scale uncertainty is reduced throughout the invariant-mass range considered, whereas the factorization scale uncertainty gets increased which is about 28\% at low $Q$ and decreases almost linearly with $Q$, reaching a minimum of about 1\% near $Q\simeq 1.3$~TeV, beyond which it rises again towards high $Q$.
To show this turnover, the $\mu_F$ panel is extended to $Q=1.5$~TeV, whereas the $\mu_R$ panel is shown up to 1~TeV, as in the other figures.

\begin{figure}[!htbp]
\centering
\includegraphics[width=\textwidth]{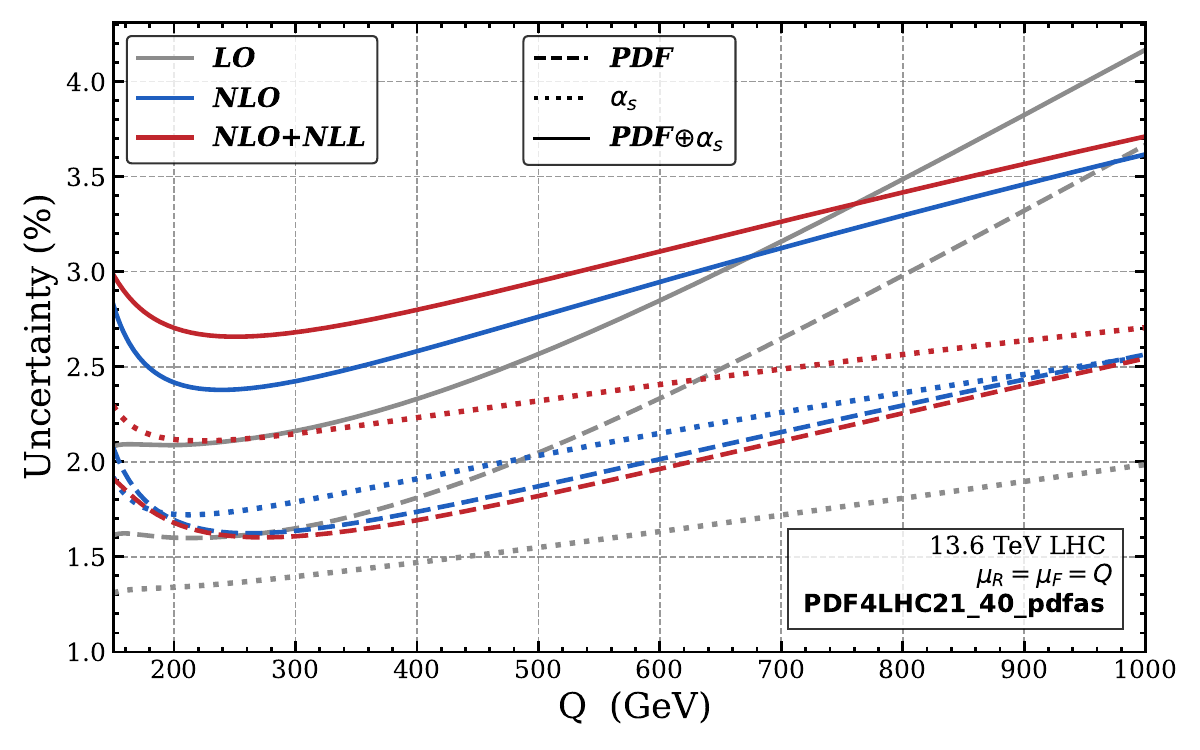}
\caption{Percentage of uncertainties, from PDF variations and $\alpha_s$ variations, 
 in the invariant mass distribution.}
\label{fig:pdf-uncertainty}
\end{figure}

Finally, we study the non-perturbative PDF uncertainties and $\alpha_s$ uncertainties across the invariant-mass spectrum. Figure~\ref{fig:pdf-uncertainty} shows the percentage uncertainties associated with PDF variations (dashed curves), $\alpha_s$ variations (dotted curves) and combined variations (solid curves) as a function of $Q$.
At LO, the PDF uncertainty rises with invariant mass, growing from $\approx 1.6\%$ at $Q=150$ GeV to over $3.6\%$ at $Q = 1000$ GeV. 
This rise reflects the expanding uncertainty in the gluon distribution function at higher partonic momentum fractions $x$. At NLO and NLO+NLL, the PDF uncertainty shows improved stability, varying mildly between $1.6\%$ and $2.0\%$ over most of the spectrum and reaching only $\approx 2.4\text{--}2.5\%$ at $1000$~GeV.
The $\alpha_s$ uncertainty increases smoothly with $Q$, from about $1.3 ~(2.5)\%$ to about $2.0 ~(2.6)\%$ at LO (NLO+NLL). 
Crucially, throughout the entire invariant-mass range $Q \in [150, 1000\text{ GeV}]$, the combined PDF and $\as$ uncertainties remain below $3.6\%$, which is roughly an order of magnitude smaller than the seven-point scale variation at NLO+NLL. 
These large scale uncertainties can only be reduced by including even higher order QCD corrections.
Finally, we note that starting from LO we have considered only five quark  flavors in the massless limit. To consider the impact of finite top-quark mass effects in our calculation, we follow the approach of Born-Improved Theory where the higher order QCD corrections obtained with five quark flavors will be used to rescale the leading order predictions obtained with six quark flavors.  
This approach was
used in the literature \cite{Grazzini:2018owa,Buonocore:2021fnj}.
For $13.6$ TeV, the production cross section with top-quark mass effects are small at NLO+NLL, where BIT result is $1882.6$~fb, compared to $1887.01$~fb
for five  flavors, a reduction of about  $0.2\%$.
The same reduction is observed at LO and across all seven scale choices, so the scale uncertainty remains unchanged.
\begin{figure}[!htbp]
  \centering
  \begin{subfigure}[t]{0.49\textwidth}
    \centering
    \includegraphics[width=\linewidth]{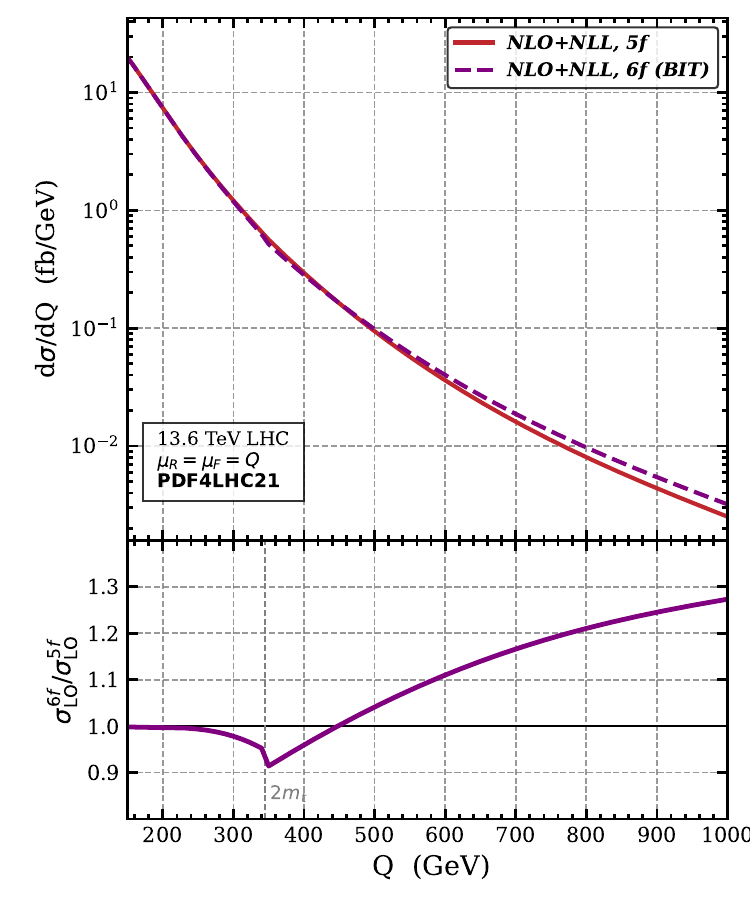}
    \caption{}
    \label{fig:BIT-distribution}
  \end{subfigure}\hfill
  \begin{subfigure}[t]{0.49\textwidth}
    \centering
    \includegraphics[width=\linewidth]{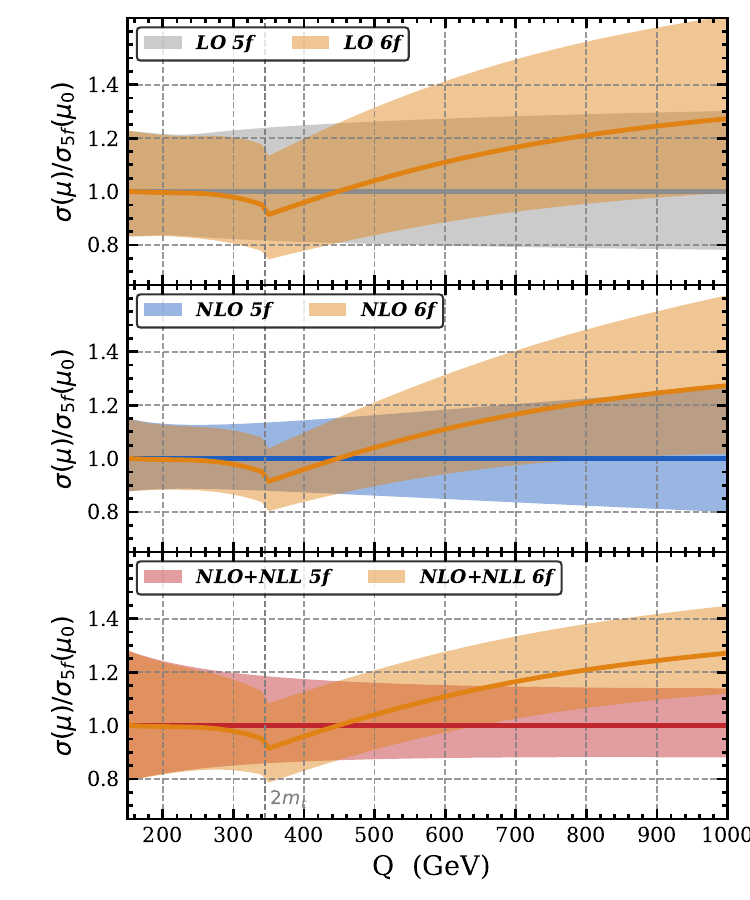}
    \caption{}
    \label{fig:BIT-scale-bands}
  \end{subfigure}
  \caption{(a) Invariant-mass distribution of the $Z\gamma$ pair at NLO+NLL accuracy
  computed with
  $n_f =5$ massless quark flavors
  and with $n_f=6$
flavors including the top quark contribution
  in BIT approach (see the text). The lower panel shows the ratio
    $\sigma^{6f}_{\rm LO}/\sigma^{5f}_{\rm LO}$.
    (b) Seven-point scale-uncertainty bands for both
    $\sigma_{5f}$
    and $\sigma_{6f}$
    at LO, NLO and NLO+NLL, normalized to the
    respective $\sigma_{5f}(Q,Q)$.}
  \label{fig:BIT-distribution-ratio}
\end{figure}
In Figure~\ref{fig:BIT-distribution} we show the invariant-mass distribution obtained at NLO+NLL accuracy for five flavors and for six flavors in the BIT approach. We observe that the two predictions are very close below the top-pair threshold ($2m_t$). Near the top-pair threshold region, the six flavor result is suppressed by about $8\%$.
A more accurate description of this region necessitates the computation of two-loop amplitudes with the full top-quark mass dependence.
 Figure~\ref{fig:BIT-scale-bands} represents the seven-point scale uncertainties obtained with six flavors normalized with five flavors.  
Above the top threshold, the top-quark mass effects are found to increase so that the cross section with six flavors is about  $27\%$ larger than the five flavor result at around  $Q=1000$ GeV. 

\section{Conclusion}\label{sec:conclusion}
In this work, we have presented the NLO QCD corrections to the loop-induced $gg\to Z\gamma$ production at hadron colliders. These results were obtained with an in-house code based on the Catani--Seymour dipole subtraction method. The two-loop virtual amplitudes are computed using the helicity amplitudes available in the literature \cite{Gehrmann:2013vga}.  
In this process, the photon is isolated using Frixione smooth-cone isolation criterion.
Further, we have extended our work to include the effect of threshold logarithms by resumming them up to NLL accuracy and matching to the fixed-order NLO result.

At the 13.6~TeV LHC, the NLO corrections are large. They increase the LO production cross section by about $83\%$.

The seven-point scale uncertainty was also found to be reduced from about $22.3\%$ at LO to $15.2\%$ at NLO. Threshold resummation at NLL increases this NLO cross section further by about $23\%$.
A similar pattern is observed for other collider energies from 7 to 14~TeV.

We have also studied the invariant-mass distribution and found that the $K$-factors increase with $Q$. The NLO (NLO+NLL) $K$-factors raise from about 
$1.7~(2.1)$ at $150$~GeV 
with $Q$ to as large as $3.6~(4.1)$ at $1000$~GeV. 
The seven-point scale uncertainty at NLO grows 
to about $25\%$ for  $1000$~GeV which reduces to about $16\%$ with the inclusion of resummation contribution at NLO+NLL accuracy.
Further, we also estimate the uncertainties due to $\alpha_s$ and the non-perturbative PDFs in our results, and we find both of them to be between $(1.7-3.0)$\% at NLO as well as at NLO+NLL accuracy, 
over the whole invariant-mass range considered here. 
The theoretical uncertainty is therefore dominated by scale variation, and reducing it further will require even higher-order QCD corrections. 

Finally, we estimate the contribution of the top quark in the loop amplitudes by presenting the invariant mass distribution obtained in the Born-Improved theory. We find that beyond $650$ GeV the top mass effects are non-negligible.
We anticipate that these results for the $Z\gamma$ production in gluon fusion will be useful in the precision studies as well as new Physics searches for the current LHC and future high energy hadron colliders.

\vskip 0.5cm

\section*{Acknowledgements}
We would like to thank Prof. V. Ravindran and Dr. Pulak Banerjee  for very useful discussions.
We acknowledge National Supercomputing Mission (NSM) for providing computing resources of ‘$\tt{PARAM \ Kamrupa}$’ at IIT Guwahati, which is implemented by C-DAC and supported by the Ministry of Electronics and Information Technology (MeitY) and Department of Science and Technology (DST), Government of India, where most of the computational work has been carried out.
C. Dey was supported in part under National Natural Science Foundation of China (NSFC) grant W2441004.

\pagebreak

\bibliographystyle{JHEP}
\bibliography{bibliography}

\end{document}